# Nickel Hexacyanoferrate Electrodes for Continuous Cation Intercalation Desalination of Brackish Water

S. Porada,[1,2] A. Shrivastava,[3] P. Bukowska,[1] P.M. Biesheuvel,[1] and K.C. Smith[4,*]

[1]Wetsus, European Centre of Excellence for Sustainable Water Technology, Oostergoweg 9, 8911 MA Leeuwarden, The Netherlands. [2]Soft Matter, Fluidics and Interfaces Group, Faculty of Science and Technology, University of Twente, Drienerlolaan 5, 7522 NB Enschede, The Netherlands. [3]Department of Materials Science and Engineering, University of Illinois at Urbana-Champaign, Urbana, IL 61801, USA. [4]Department of Mechanical Science and Engineering, Computational Science and Engineering Program, and Beckman Institute for Advanced Study, University of Illinois at Urbana-Champaign, Urbana, IL 61801, USA.

*Corresponding author's email: kcsmith@illinois.edu

## Abstract

Using porous electrodes containing nickel hexacyanoferrate (NiHCF) nanoparticles, we construct and test a device for capacitive deionization in a two flow-channel device where the intercalation electrodes are in direct contact with an anion-exchange membrane. Upon negatively charging NiHCF, cations intercalate into it and the water in its vicinity is desalinated; at the same time water in the opposing electrode becomes more saline upon positively charging the NiHCF in that electrode. In a cyclic process of charge and discharge, fresh water is continuously produced, alternating between the two channels in sync with the direction of applied current. We present proof-of-principle experiments of this technology for single salt solutions, where we analyze various levels of current and cycle durations. We analyze salt removal rate and energy consumption. In desalination experiments with salt mixtures we find a threefold enhancement for $K^+$ over $Na^+$-adsorption, which shows the potential of NiHCF intercalation electrodes for selective ion separation from mixed ionic solutions.

## 1.1 Introduction

Capacitive deionization (CDI) makes use of porous electrodes to store salt ions. One commonly used material is porous carbon [1]. Porous carbon electrodes can be made of carbon nanotubes, graphene, activated carbon powder, etc., and processed into porous, ion- and electron-conducting, thin electrode films, suspensions, or fluidized beds [2]. CDI with carbons is a promising method, but to reach a certain equilibrium salt adsorption capacity (eqSAC; a typical number being of the order of 5-15 mg/g, referring to mass of NaCl removed, per total mass of carbon in a two-electrode cell, measured at a standard charging voltage of $V_{ch}$=1.2 V with discharge at 0 V), the energy input is not insignificant [3–5], while the current efficiency $\lambda$ (quantifying the fraction of current input that results in salt adsorption) can be well below unity, implying that in the charging process not only counterions adsorb but also coions desorb from the electrode [6]. In CDI with membranes, or using improved charging schemes, $\lambda$ can be close to unity [5].

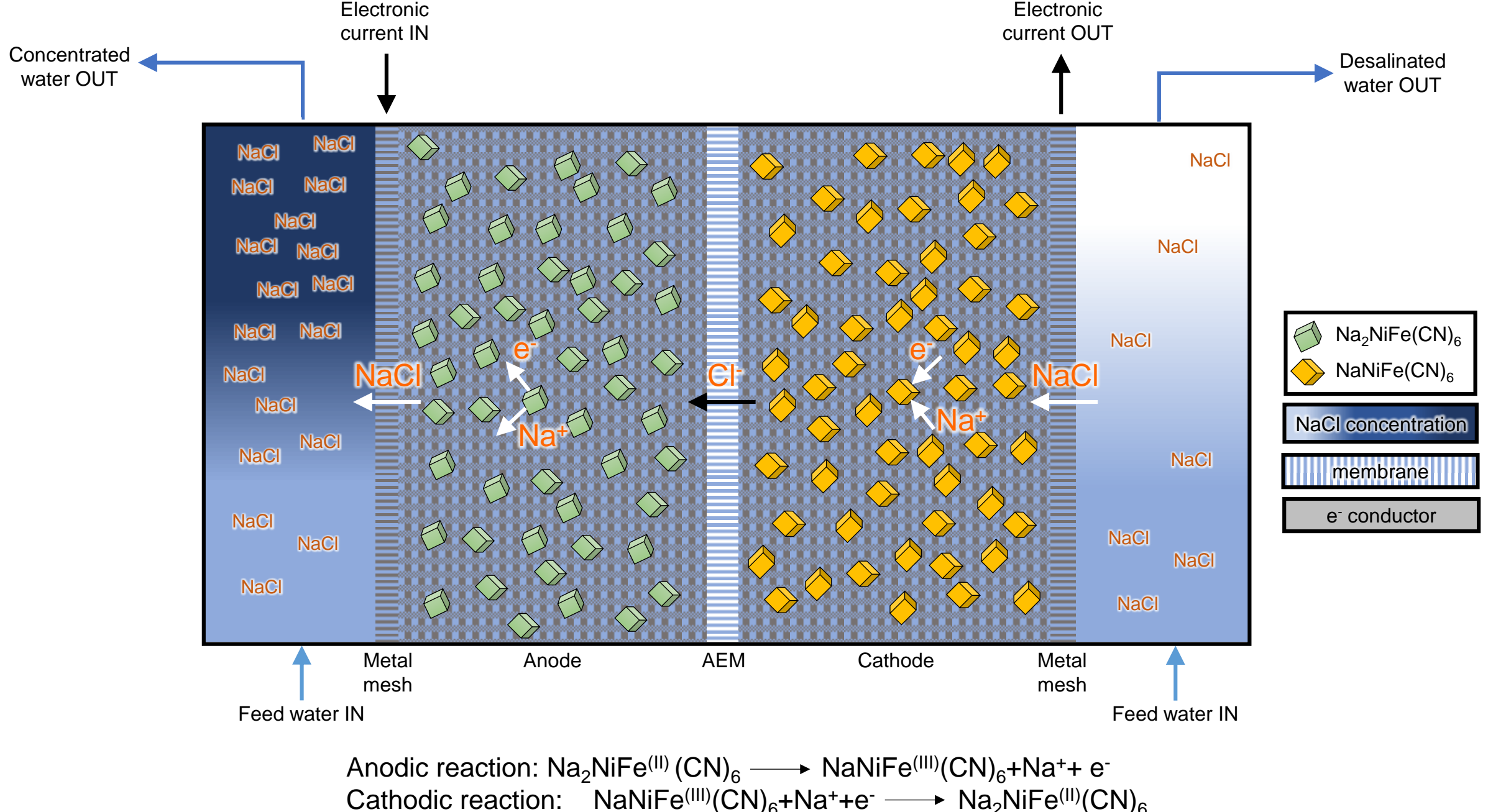

**Figure 1.** Schematic diagram of the present CID cell with a single anion exchange membrane (AEM) and two adjacent and identical NiHCF IHC electrodes. Flow channels and current collectors are on both sides of the cell. When current runs from left to right cations (in this case, $Na^+$ ions) de-intercalate from $Na_2NiFe(CN)_6$ IHC particles while $Cl^-$ ions migrate through the membrane (to the left), the net effect of which is to produce water of increased salinity on the left-hand side. At the same time, on the right-hand side, $Na^+$ ions intercalate into $NaNiFe(CN)_6$, and desalinated water is produced. After some time, the maximum amount of intercalated $Na^+$ is reached in the right-hand electrode and the current direction must be reversed (not shown).

Intercalation materials are a new type of electrode material which are recently receiving much attention [7–9]. In intercalation electrodes ions are stored in the crystallographic sites of an intercalation host compound (IHC). Porous electrode films can be prepared from small (order of 10 micron) IHC particles, see Fig. 1. Water desalination using IHCs potentially has the advantage that to reach a certain SAC a much lower voltage and energy is needed than with CDI based on carbons. This is because from a rather low to a rather high charge, the change in electrode potential with electrode charge is much lower with IHCs than with carbons, i.e., the differential capacitance of IHCs (a number with unit F/g) can be much higher than for EDL formation in carbons at the same charge as IHCs. Also, IHCs have the potential to selectively remove one ion out of a multi-ion mixture with other ions of the same valence and charge [10,11].

The use of IHCs for water desalination has been reported previously using several novel cell architectures. The “Desalination Battery” (DB), a cell consisting of one $Na_2Mn_5O_{10}$ (NMO) electrode (and later with $Na_{0.44}MnO_2$ [11]) and one Ag/AgCl electrode (and recently with BiOCl [12,13]), was used for water desalination by adsorption of cations within NMO and adsorption of $Cl^-$ ions by conversion of Ag to AgCl [8]. In “Hybrid CDI” (HCDI) [7], an IHC electrode (e.g., with NMO [7], $Na_2FeP_2O_7$ [14], $MoS_2$ [15], or $NaFe_2(CN)_6$ [16]) was combined with a carbon electrode for anion adsorption. In both of these desalination devices a single cation-intercalating electrode was paired with a different electrode that padsorbs anions, that has either economic (due to costly Ag electrodes [8]) or capacity (due to the EDL adsorption mechanism [7]) limitations. Smith and Dmello (SD) [9,17] proposed desalination with Na-ion IHC electrodes having identical chemical composition, originally referred to as “Na-Ion Desalination (NID)” [9]. Presently, we refer to this technology as Cation Intercalation Desalination (CID) to emphasize the charge adsorption mechanism by which desalination is achieved. While the use of the same IHC in both electrodes is not common for battery use, SD showed that desalination is possible, in theory, with electrodes of identical composition but with a different degree of intercalation. To achieve this, an innovative cell design was proposed where porous IHC electrode films are placed on either side of a separator layer with feed water in a “flow-through” mode directed through the electrode, along the separator layer. SD showed that for an IHC that is cation-adsorbing, the separator layer must be cation-blocking to achieve high salt removal, i.e., an anion-exchange membrane (AEM) must be used. In this way a CID device operates where, during one half of a charging-discharging cycle, one channel produces desalted water and the other brine, and, during the other half of the cycle, the situation is reversed. This concept is experimentally demonstrated in our precent work (published as a preprint in ref. [18]). Recent modeling [19] has shown that the CID cell concept can be extended to include multiple flow channels between NiHCF electrodes, producing an ED stack that uses intercalation electrodes rather than the gas evolution reactions typically used in ED, as experimentally demonstrated in ref. [20]. Recently an asymmetric configuration of Prussian Blue analogue electrodes was used without flow to desalinate seawater [21]. Presently we employ symmetric NiHCF electrodes with brackish solution flowing through porous spacers, enabling continuous production of desalted water.

The recent development of IHCs for aqueous rechargeable Na-ion batteries (ARSBs) has been stimulated by their low cost compared with non-aqueous Li-ion systems, and by the inherent safety of water-based electrolytes [22]. By leveraging the CID device concept these developments can be applied to capacitive deionization. Also, the plethora of Na-ion IHCs developed as cathodes for ARSBs can be applied as both cathode and anode in CID, eliminating the need to search for low-potential IHCs, which, for ARSBs, are less numerous compared to high-potential IHCs. Theoretical predictions of CID performance were originally obtained [9] using symmetric cells with either NMO, a common ASRB cathode [23–25], or $NaTi_2(PO_4)_3$ (NTP), a common ASRB anode [24,26,27]. These respective materials have high specific capacities of 45 mAh/g-NMO [23] and 120 mAh/g-NTP [24], but other IHCs exist that also show promise for use in CID. Among the various cathode compounds developed, nanoparticulate Prussian Blue Analogues (PBAs) stand apart for their high cycle life and facile cation intercalation kinetics that have enabled their demonstrated use in $Na^+$ [27–30], $K^+$ [28,29], $Mg^{2+}$ [31], $Ca^{2+}$ [31,32], and $Zn^{2+}$ [33] ion batteries. Here, we use a particular PBA, nickel hexacyanoferrate (NiHCF), to exchange $Na^+$ ions reversibly with saline solution flowing through a CID cell. Its average reduction potential of ~0.6 V vs. SHE for Na-ion intercalation [27,28], though modest as a cathode in ARSBs, is 200 mV below the $O_2$ evolution potential at neutral pH. As a result, minimal electrolyte decomposition is expected even at large overpotentials applied to CID cells using these IHCs. Also, Smith recently predicted that NiHCF has sufficient capacity to desalinate seawater [19]. Here, we demonstrate the CID concept experimentally using NiHCF electrodes with brackish-water level NaCl/KCl influent. Based on the results in our present work, we calculate that this material can potentially achieve eqSACs of ~40 mg/g-electrode (assuming 59 mAh/g-NiHCF capacity [27,28]; current efficiency 92%; charge capacity utilization 85%; IHC 80 wt% of electrode mass; parameters to be explained later on).

Technologies bearing similarity to CID (in terms of either the materials or cell construction used or of device functionality) have been reported previously in the literature, but the present work is the first experimental demonstration bringing together two porous cation-intercalation electrodes separated by an AEM to desalinate water continuously from a feed-water stream. In a technology called "electrically (or, electrochemically) switched ion-exchange" (ESIX) electrodes are used to achieve a separation between different cations ($Na^+$, $K^+$, $Cs^+$), first only with IHC used as cathode, with NiHCF deposited on a Ni-electrode [34] (see also Ref. [35] for CuHCF coated from a water based solution directly on the electrode surface), and later including an AEM and a symmetric cell with two identical NiHCF electrodes prepared by deposition on carbon felt [36]. Distinct from earlier literature, we use symmetric NiHCF intercalation electrodes to desalinate NaCl solution representative of brackish water using an AEM to separate these electrodes, and we determine desalination-relevant performance metrics, in contrast with previous experiments focused on selective $Cs^+$ removal [34–36]. We also prepare self-supporting porous electrodes of around 250 μm thickness containing macroporosity for ion transport and carbon additives for electronic transport. We note that similar architectures have been used with redox-active polypyrrole polymer electrodes, rather than IHCs, separated by an AEM have been used to soften water through the removal of $Ca^{2+}$ [37]. Also, Ag/AgCl conversion electrodes have been used to desalinate seawater with coulometric precision (one electron adsorbs one salt molecule) when separated by a cation-exchange membrane [38]. In the field of osmotic power (energy harvesting from a water salinity difference), recent work with identical cation-intercalation electrodes was reported using a cell with a non-selective membrane [39,40]. Different from that work, we explore the selective removal of these ions under continuous flow and using a CID cell configuration with symmetric electrodes separated by an AEM, and we measure time-varying effluent concentrations of these ions. Additionally, superlattice materials (e.g., $TiO_xC_y$ or MXene [41]) have been reported to desalinate water via intercalation of both cations and anions in a membraneless configuration.

In this work we test and demonstrate continuous desalination within a CID cell for the first time experimentally. Here, desalination is achieved by using NiHCF nanoparticles to intercalate $Na^+$ ions from NaCl solution. We synthesize Na-rich NiHCF using precipitation chemistry and fabricate porous electrode films from them. Prior to CID cell testing electrochemical titration is used to characterize the equilibrium relationship between potential and charge in a single electrode film using non-flowing, concentrated $Na_2SO_4$ electrolyte. These porous electrode films are then integrated within a CID cell adjacent to an AEM (Fig. 1). Feed water is flowed behind the porous electrodes through flow channels adjoined to each of the cell's current collectors, enabling continuous desalination of feed water. The dynamic, cyclic operation of this CID cell is subsequently characterized at a salinity that is representative of brackish water. A systematic study of desalination performance is then conducted by varying cycle time with the same current in each case, demonstrating the range of charge, salt adsorption capacity, and energy consumption achievable using NiHCF-based electrodes in a CID cell. Finally, we also show experimental results for desalination in $Na^+$/$K^+$-mixtures and the uptake of each ion by the IHC.

## 2. Experimental Methods

### 2.1 Synthesis and Material Characterization

Disodium nickel hexacyanoferrate was synthesized by a solution/precipitation reaction with aqueous reagents using the following synthesis reaction: $NiCl_2 + Na_4Fe(CN)_6 \rightarrow Na_2NiFe(CN)_6 + 2NaCl$. A previous recipe used to synthesize Na-rich NiHCF [27] was employed with several modifications by incorporating findings from a recipe used to synthesize Na-rich manganese hexacyanoferrate [42]. In this procedure, 20 mM of $Na_4Fe(CN)_6 \cdot 10H_2O$ (Sigma Aldrich) and 2.40 M of NaCl (Sigma Aldrich) were dissolved in a mixture of pure water with 25 wt% pure ethanol (compared to pure water). Subsequently, 40 mM of aqueous solution of $NiCl_2 \cdot 6H_2O$ (Alfa Aesar) was added dropwise to the mixture of $Na_4Fe(CN)_6$, NaCl, water and ethanol. The solid NiHCF precipitate was filtered, washed with pure water, dried in a vacuum oven at 50°C to remove residual moisture, and ground using a ball mill.

Elemental analysis was conducted with NiHCF decomposed into elemental species in acid using a PerkinElmer 2400 Series II CHN/O Elemental Analyzer and a PerkinElmer 2000DV ICP-OES instrument. X-ray diffraction was conducted using a Bruker D8 Venture instrument with Cu-Kα radiation (λ=1.54 Å, 0.02˚ angular resolution, and 3˚ < 2Θ < 100˚). Both samples showed slight peak broadening indicative of their nanocrystalline microstructure, and several high-angle peaks were not apparent due to background noise. Also, it should be noted that fluorescent interaction between Fe and X-Ray radiation emitted from Cu can produce mild inaccuracy in the measured intensity. $N_2$ adsorption measurements were performed at -196°C using a TriStar 3000 gas adsorption analyzer (Micromeritics).

SEM samples were deposited and pressed gently onto carbon tape attached to an aluminum sample holder. To minimize charging the samples were coated with approximately 4 nm of gold-palladium film using a Denton (Moorestown NJ) Desk-II turbo sputter coater, and imaged at 5 kV in HiVac mode using a Quanta 450 FEG environmental scanning electron microscope (FEI Company, Hillsboro OR). TEM samples were prepared by dispersion in deionized water by sonication in a water bath for approximately 5 minutes, followed by vibration using a Fisher S56 Miniroto touch shaker. Dried droplets were then glow-discharged on wax film and subsequently transferred to carbon-stabilized Formvar-coated TEM grids (cat. No. 01811; Ted Pella, Inc., Redding CA). TEM images were obtained using a Philips/FEI CM200 instrument at 160 and 200 kV.

2.3 Electrode Preparation and Characterization

80 wt% of the NiHCF powder was mixed with 10 wt% carbon black (Vulcan XC72R, Cabot Corp., Boston, MA) and 10 wt% polytetrafluoroethylene binder (PTFE, 60 wt% solution in water from Sigma Aldrich, USA) and pure ethanol to obtain an homogeneous slurry. The final electrodes were calendered using a rolling machine (MTI HR01, MIT Corp.) to produce a thickness of approximately 250 μm. Prior to CID experiments, electrodes were cut into 6x6 cm squares and dried in a vacuum oven at 50°C to remove residual moisture, which is necessary to measure electrode dry mass.

Electrochemical characterization of the NiHCF material was performed by constructing a three-electrode cell with the working electrode containing 0.88 g of electrode in 1 M $Na_2SO_4$, cycled against a titanium mesh electrode coated with Ir/Ru (Magneto Special Anodes B.V., the Netherlands) as counter electrode and Ag/AgCl as a reference electrode placed in the vicinity of the working electrode, as reported in Fig. 2. Titration data in Fig. S5 for 0.1 M $Na_2SO_4$ and 0.1 M $K_2SO_4$ use an electrode with $M_{el}$=0.77 g; the same electrode is also used in Fig. 5. In "bursts" of 5 C (at a current of 60 mA/g-NiHCF) the working electrode was charged, after which a "rest period" ("OCV" for "open circuit voltage") of 60 s follows in which the potential relaxes (see inset in Fig 2A). We continued this process until the electrode potential diverged strongly, implying full (de-)intercalation of $Na^+$ ions was reached. The electrode potential at the end of each OCV period was taken as equilibrium potential (versus Ag/AgCl) and plotted as a data point in Fig 2B.

2.4 Desalination Cell Design and Testing

The experimental CID flow cell was constructed in the following way. Two end plates on the outer sides of the cell were used to sandwich meshed current collectors, NiHCF electrodes, and an AEM. In this work we used a commercial AMX Neosepta membrane with a thickness of 140 μm. Meshed current collector also served as a spacer material to allow feed water to flow along the NiHCF electrode surface. For the experiments reported in Figs. 3, 4, 5 and S3, the cell contained an electrode of $M_{el}$=0.99 g (of which ~0.80 g is NiHCF) on each side, and the membrane area available for ion transfer was 36 $cm^2$. Desalination experiments were performed at a constant current of 5 or 10 mA, and a water flow rate of 10 mL/min directed into each channel (4.7 mL/min per channel in Fig. 5), producing a 50% water recovery rate (defined as the ratio of desalinated water volume to total influent volume). Higher water recovery levels can be achieved by reducing flow rate in the concentrating flow channel, as predicted previously [9]. Constant current was applied until a certain charge was transferred, as shown in Figs. 3 and 4 per gram of NiHCF material in one electrode. Subsequently, the current direction was reversed. After two or three similar cycles with the same total charge transfer, the charge transfer was reduced stepwise. Because all experiments in Fig. 3 were done at the same current density (which can only switch sign), the cycles become shorter in duration. For the mixture experiments reported in Fig. 5, the electrode mass is $M_{el}$=0.77 g. Ion concentration during the mixture experiments was measured using an on-line method in which part of the effluent water flows directly to an inductively coupled plasma optical emission spectroscopy (ICP-OES) instrument (for more details about this method see Ref. [43]).

## 3. Results and Discussion

### 3.1 Structural and Physical Characterization

NiHCF has previously been synthesized for Na-ion batteries in both Na-deficient [28] and Na-rich [27] forms, the latter form enabling batteries to be assembled and charged without pre-sodiation when it is used as a positive electrode. In this work, we synthesize NiHCF in a Na-rich state, with $Na_2NiFe(CN)_6$ being the ideal composition [27]. Synthesis of Na-rich NiHCF (with $Na_4Fe(CN)_6$ precursor) eliminates introduction of additional cationic species that are present in the typical synthesis of Na-deficient NiHCF, which, for example, used $K_3Fe(CN)_6$ precursor in ref. [28] and, thus, resulted in $K^+$ within the synthesized NiHCF structure. The present Na-rich NiHCF material is later converted to intermediate compositions via electrochemical oxidation in a three-electrode cell. NiHCF particles undergo nucleation and growth as the precursors are combined, and the manner in which this is done will affect the final particle size distribution as well as the crystallinity of the synthesized particles. We use a solution/precipitation method with consistent reaction conditions to synthesize NiHCF (see Experimental Methods).

Based on ex situ elemental analysis Na-rich (as-prepared) NiHCF samples were found to have a composition of $Na_{1.20}Ni_{1.28}Fe_{1.22}C_6N_{5.75}\cdot4.41H_2O$ (see Methods and Fig. S1). This composition includes significantly less Na per formula unit CN than that of the theoretical composition ($Na_2NiFe(CN)_6$), also with a substantial amount of hydrated $H_2O$ within its lattice. In light of these factors, the presence of $H_2O$ trapped within the interstitial sites of the NiHCF structure may play a role in limiting the incorporation of $Na^+$ ions. Despite their non-ideal composition, we note that the present NiHCF possesses more than double the degree of alkali cation intercalation and similar nickel stoichiometry per CN formula unit compared with those of $Na^+$/$K^+$-deficient NiHCF synthesis ($K_{0.6}Ni_{1.2}Fe(CN)_6\cdot3.6H_2O$, ref. [28]). These observations demonstrate that we have successfully synthesized Na-rich NiHCF. The crystal structure of as-synthesized NiHCF was confirmed by X-ray diffraction (see Methods) to belong to the Fm3m space group with a lattice parameter of 10.25 Å, in good agreement with literature [28]. Peak positions match well with previous reports in literature for Prussian Blue [44] and NiHCF [27,28] (see Fig. S1 in Suppl. Inf., SI). The crystallinity and physical morphology of NiHCF particles were analyzed using transmission (TEM) and scanning electron microscopy (SEM, Fig. S1). TEM images show nanocrystallites between 2 to 7 nm, with mean crystallite size of approximately 4 nm. Nanocrystallite shapes appear as rounded cubes (in contrast with previous reports of highly faceted PBA nanoparticles [45]), likely as a result of crystal defects introduced during nanocrystal growth. SEM images of the NiHCF particles showed a range of primary and secondary particle sizes. While secondary particles were observed with diameters on the order of 10 µm, detailed examination at higher magnification reveals the presence of nanoparticles with diameters on the order of 100 nm, consistent with prior reports on NiHCF [27,28].

The nanoparticulate morphology of NiHCF was further studied by $N_2$ gas adsorption. This measurement was also used to assess the size distribution of pores in the electrodes (see Methods). As shown in Fig. S1, up to a pore size of 30 nm, porosity analysis of synthesized NiHCF powder resulted in a pore volume of 0.028 mL/g, a BET area of 15 $m^2$/g, and a surface area using non-linear NLDFT theory of 12 $m^2$/g [46]. Assuming a density for NiHCF of 2.0 g/mL (based on its ideal stoichiometry and XRD analysis), we find a corresponding pore volume per unit NiHCF volume of 5.6%, revealing ultra-low microporosity indicative of nanoparticle aggregation. Accordingly, this powder was ball milled to separate nanoparticle aggregates (see Methods). Electrodes were subsequently prepared using a procedure similar to that reported in ref. [47], combining 80 wt% NiHCF with 10 wt% PTFE binder and 10 wt% conductive carbon black, after which the electrodes were calendered to increase electronic conductivity (see Methods). After this process the BET area of the electrode was 103 $m^2$/g, and pore volume (in the range to 30 nm pore size) was 0.12 mL/g or 29 vol% (assuming 2.0 g/mL density, as before), confirming the separation of nanoparticulate aggregates that is necessary for high electrochemical activity. SEM characterization of the electrode showed a homogeneous microstructure consisting of NiHCF, carbon black, and macroporosity (see SI). The macropores within the electrode (which, during cell operation, is filled with saline water) are estimated to have a porosity of 40 to 50 vol%.

### 3.2 Single Electrode Characterization

In the context of desalination of an NaCl solution by CID, charging of an electrode translates to salt removal, as $Na^+$ ions are the primary intercalant species storing charge within the electrodes, and, therefore, understanding the cycling characteristics of each electrode is essential to constructing a high-performance CID cell. To illustrate this principle, consider a dimensionless number between 0 and 1 denoting the fraction of cation-filled sites within the IHC, hereafter referred to as the intercalation degree θ. For an electrode with θ=0 the IHC contains no exchangeable intercalant, while at θ=1 the IHC is saturated with intercalant. To convert from intercalation degree θ, for which 0<θ<1, to electrode charge $Q$ in mAh/g, we multiply by the maximum charge $Q_{max}$ at full intercalation, i.e., $Q=Q_{max}\cdot\theta$. For single electrode testing, equivalent spans of these two scales are shown in Fig. 2B. In desalination with Na-ion IHCs (such as NiHCF), the intercalation degree within opposing electrodes evolves in a symmetrical fashion, switching in the fraction of intercalated Na, θ, between two states of θ (see ref. [9,48]). During one half of the cycle the composition in one electrode changes from $\theta_1$ to $\theta_2$ ($\theta_2>\theta_1$), while in the other electrode, simultaneously, the composition changes from $\theta_4$ to $\theta_3$ ($\theta_4>\theta_3$). Such operation is referred to as the “rocking-chair” cycling mechanism utilized commonly in Li-ion [49] and many other rechargeable batteries using IHCs. The importance of this principle to the CID concept is exactly this: cations are intercalated from the electrolyte in one electrode, while at identical rate cations are deintercalated into the electrolyte in the opposing electrode [9]. When the IHCs in both electrodes are identical and are of equal mass, this action produces an equal and opposite change in composition, i.e., $\theta_2-\theta_1=\theta_4-\theta_3$. Thus, based on the use of IHCs an innovative cell design is proposed with only one ion-exchange membrane and two chemically identical IHC electrodes.

NiHCF electrodes were characterized in 1 M $Na_2SO_4$ solution (often used in ARSBs) using the galvanostatic intermittent titration technique (GITT, see Methods). Typically, when IHCs are used for energy storage in batteries, the principal metric of consideration is the maximum electrode charge in mAh/g. The equilibrium potential versus electrode charge in Fig. 2B, extracted from GITT, shows a maximum charge of 59 mAh/g (similar to previous reports for NiHCF [27,28]) with potential varying in a roughly linear manner between an electrode charge of 5 and 55 mAh/g (per gram NiHCF in one electrode). From this characterization we also find that the relationship between equilibrium potential and the intercalation degree within NiHCF particles can be correlated by the Frumkin isotherm [50],

$$E=E_{ref}-RT/F^{*}\{\ln(\theta/(1-\theta))-\ln(c_{Na,\infty}/c_0)\}-g\cdot(\theta-½), \quad (1)$$

where at room temperature $RT/F$=25.6 mV and the interaction parameter $g$ is positive when intercalated Na-ions in the IHC repel one another, and therefore do not phase-separate into Na-rich and Na-deficient regions (which is predicted when $g$ smaller than -100 mV). Here, $c_{Na,\infty}$ is the $Na^+$ concentration within the electrolyte surrounding IHC particles, and $c_0$=1 M is a reference concentration. A best fit to the data is obtained using $E_{ref}$=425 mV vs. Ag/AgCl and $g$=+90 mV. Such a correlation is also an essential input for porous-electrode modeling of Na-ion batteries and CID, which can describe the dynamics of ions within electrodes during operation [9,19].

We now compare NiHCF electrodes to EDL-based carbon electrodes on the basis of material-intrinsic desalination metrics. From the data in Fig. 2, we calculate for IHC electrodes the differential capacitance, *C*, as the derivative -d*Q*/d*E*, and we plot these data in Figure S2 as a function of charge and as a function of electrode potential. At an optimum electrode charge, capacitance is as high as 1000 F/g, to drop to lower values away from this point (e.g., capacitance drops by two when the charge is changed by about 25 mAh/g). These values are clearly higher than for carbon electrodes which typically have a capacitance not higher than 100 F/g (defined per single electrode voltage and single electrode mass). This difference in capacitance implies that, to store the same charge, a much lower voltage may be needed for intercalation electrodes than for carbon electrodes. As an example, to achieve the same salt adsorption capacity, with *C* ten-fold higher and assuming the same current efficiency, we require a final cell voltage that is ten-fold lower, and thus, with the energy proportional to $C \cdot V^2$, the energy input can be ten-fold lower as well. Note, this analysis only holds in a window of charge (or potential) where *C* is fairly constant, and neglects any mass transfer limitations and Ohmic losses.

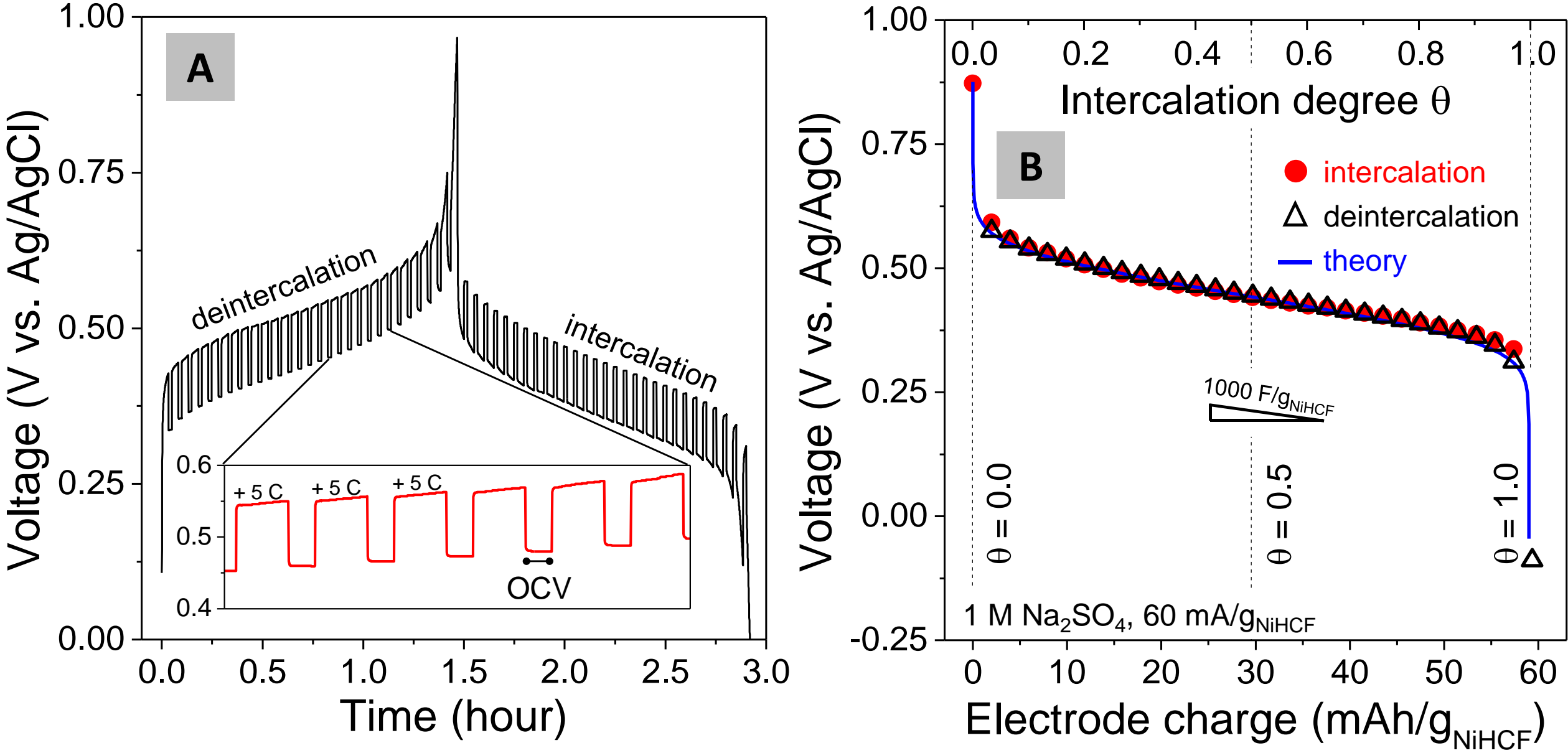

**Figure 2.** Electrochemical titration of a single NiHCF IHC electrode in 1 M $Na_2SO_4$. In this experiment, a constant current of 60 mA/g-NiHCF is applied for 118 s, after which current is set to zero for 60 s. Around 1.5 hr, the current direction is reversed. From the time-dependent trace of voltage (panel A), we take the voltage at the end of the “rest”-period and construct panel B, which shows the equilibrium potential vs charge of the NiHCF-material, with excellent reversibility between de- and intercalation of $Na^+$. The Frumkin-equation fits data according to Eq. (1) well.

### 3.3 Desalination Dynamics with Brackish NaCl Solution

Desalination experiments, as reported in Fig. 3, are performed at a salt concentration in the feed water of 20 mM and a current density of 2.8 $A/m^2$ (see Methods). A constant current (CC) mode was used because (1) simulation has predicted that such operation produces nearly constant effluent concentration after an initial transient period [9,19] and (2) recent analysis of CDI cycling modes has shown that CC mode can be more energy efficient than the constant voltage mode [4]. This particular current density is small in comparison with typical CDI experiments, but large cell polarization at high currents prevented further increases in current density at the present feed water salinity level. The large cell voltages that developed at higher currents may be due to an ion transport rate limitation in the cell. Specifically, $Na^+$ ions must transport between the porous electrode and the flow channel, and $Cl^-$ ions must transport between the porous electrodes on opposing sides of the cell through the membrane. Thus, the microstructure and thickness of the porous electrodes, as well as those of the flow channels, can be optimized to enable efficient operation at higher rates, as has been predicted previously [9]. Furthermore, the arrangement of flow channels behind the electrodes could limit achievable polarization levels, and alternative flow configurations could reduce energy consumption as recent modeling suggests [19]. In addition, because NiHCF is a type of metal-organic framework whose $CN^-$ ligands are poorly conducting, electronic conduction is expected to be slow within these particles. As such, it is possible that the inhomogeneous distribution of conductive carbon black limits transmission of electrons to NiHCF particles. Solid-state diffusion of $Na^+$ ions within NiHCF particles may also have played a role, depending on the degree of size dispersity among them. Membrane resistance is not expected to affect performance substantially due to the high concentration of membrane fixed charge in comparison with influent salinity. The exact reason for the high resistance remains to be identified and solved.

To characterize desalination dynamics, we measure the time variation of effluent salinity and cell voltage in cycles of different duration but at the same current. Data in Fig. 3 were obtained in cells where initially one electrode is almost completely de-intercalated and the other almost fully intercalated following the approach proposed by SD [9]. Here, the total charge transferred between electrodes was reduced from 50.0 to 11.1 mAh/g-NiHCF (defined per mass of NiHCF in one electrode) in steps of 5.56 mAh/g every 2-3 cycles (Fig 3 shows results of one cycle for each condition). In SI we present results of a different experiment where the cell starts with two electrodes that initially have a similar intercalation degree $\theta$ (in the range 0.4-0.6). Here, the total charge transferred during each cycle was varied in a similar manner to that of the first routine, but in increasing order from 11.1 to 50.0 mAh/g-NiHCF. Results of this experiment match very closely with those presented in Fig. 3. In Fig. 3, panel A shows the effluent salinity (minus inflow salinity) during one cycle, as measured in the effluent of one of the two channels. It is expected that the other channel has exactly opposite ("symmetric") behavior, i.e., a salinity-versus-time trace that is the mirror image (reflection about the horizontal dashed line) of the one measured, consistent with previous predictions for CID [9,19]. In a slightly different experiment, we did measure the effluent salt concentration from both channels, and, as expected, the two traces for salt effluent concentration versus time behave in a perfectly anti-parallel manner (see Fig. S4 in SI).

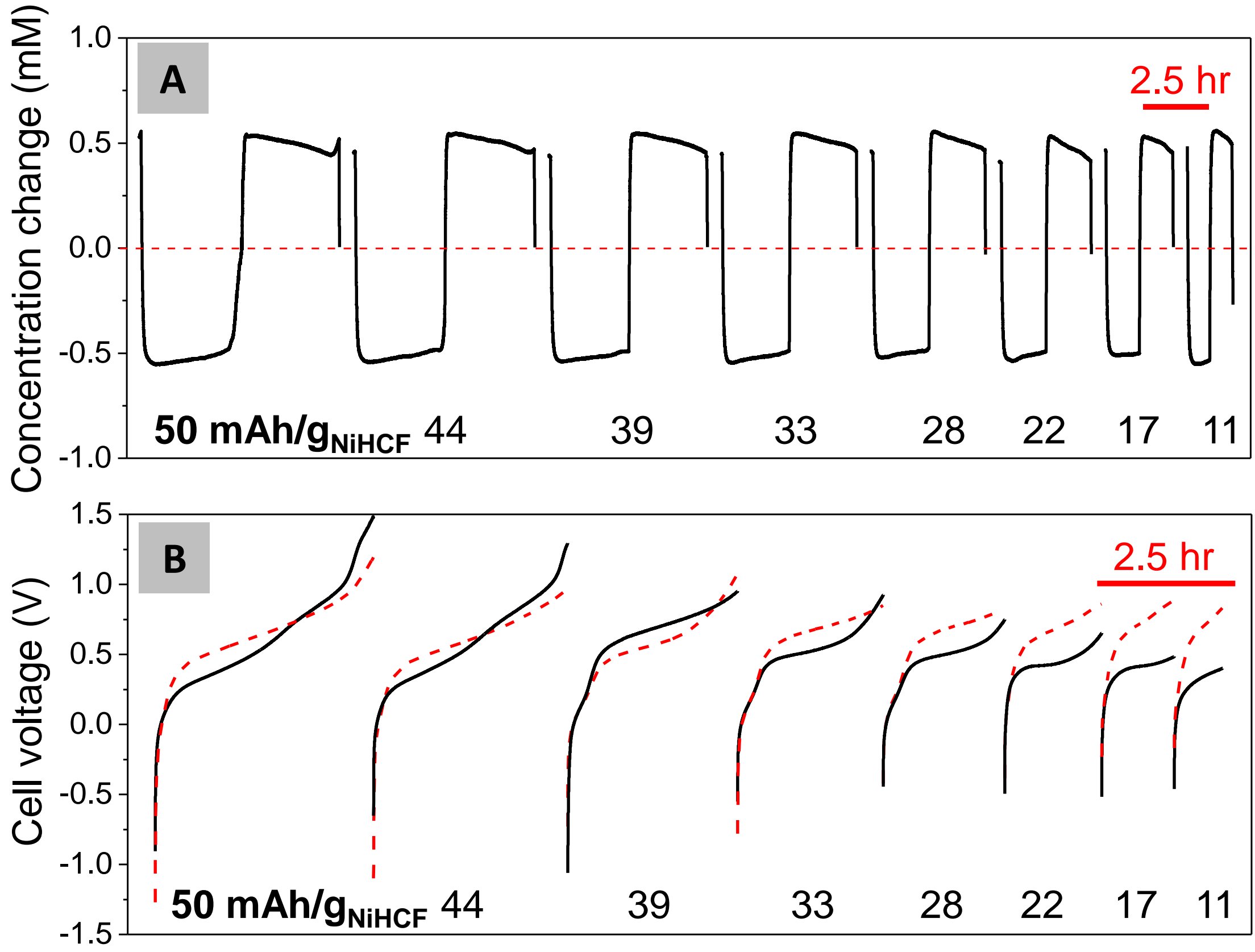

**Figure 3.** Desalination cycles using NiHCF IHC electrodes at a constant current of 2.8 A/m$^2$ in 20 mM NaCl, for sets of cycles where the charging time is reduced after each 2 or 3 cycles, starting at a situation where one electrode is almost completely intercalated with $Na^+$, and the other is in an almost completely deintercalated state. Panel A presents one cycle from each set of 2 or 3 similar cycles, showing both desalination (concentration change < 0) and salt release (concentration change > 0). Note that here only results for the effluent salinity from one of the two channels is shown. Here, the desalinated effluent concentration has been subtracted from the salinity of the influent feed water and is therefore shown as a "concentration change," where the red-dashed line indicates the influent salinity level (20 mM NaCl). In panel B profiles of cell voltage versus time are shown during the same desalination cycle. Here, to show the symmetry in the process, the voltage trace during the second half of the cycle is vertically switched and moved left (red-dashed lines), to overlap with the voltage trace in the earlier half-cycle (black-solid lines).

In Fig. 3, panel A shows that effluent salinity change (i.e., the difference between effluent and inlet salinity levels) is fairly constant, in line with the expectation for operation at constant current. With decreasing charge transfer, the half-cycle time of charging and discharging (desalination and salt release) decreases, from around 4.0 hr in the longest experiments (i.e., for a charge of 50 mAh/g, defined per gram NiHCF in a single electrode) to 0.9 hr at 11.1 mAh/g-NiHCF, but for all cycles the influent/effluent salinity change is constant. The cell voltage during charge and discharge is plotted in panel B. Here, to show overlap between the two half-cycles, the voltage-time trace during one half-cycle is vertically mirrored and shifted left, to overlap with the other voltage-time trace. In general we observe excellent agreement, and thus symmetric behavior in the two halfs of the cycle, as expected. Some differences are observed between the two traces for the lower values of charge transfer, for which we do not have an explanation. Otherwise, the two voltage-time traces overlap well. Fig. 3B shows how, at the start of each half-cycle after the reversal of current direction, the cell voltage rises rapidly in time. Such a voltage spike can typically be ascribed to the effect of a linear resistance (e.g. in wires or across a separator) and is thus linearly dependent on current. This effect is indeed largely responsible for the observed spike, but cannot explain it entirely, as its magnitude is seen to depend on the total charge in a cycle, even though the applied current is always the same. Thus, other aspects related to the state-of-charge of the NiHCF IHC must also have played a role. NiHCF electrodes were charged and discharged through more than 50 cycles with no observed degradation in performance during the various experiments conducted, and much greater cycle life is expected considering that NiHCF has previously shown sustained capacity in Na-ion batteries in excess of 4000 cycles [28].

In general, the results presented in Fig. 3 are consistent with our expectation that cell voltage gradually increases within a half-cycle, and thus for longer cycles it reaches a higher value. We also observe that – neglecting the short initial periods of negative cell voltage – in all instances the cell responds with a positive voltage when a charging current is applied (i.e., charging needs energy input). Furthermore, in none of the cycles do we see a linear increase in cell voltage with time (or, equivalently, with charge), as would be expected on the basis of the linear behavior of the “plateau” region of the voltage-charge curve in Fig. 2B. This deviation is linked to rate limitations mentioned before. This conclusion is supported by a comparison between the large cell voltage observed, to the voltage expected based on equilibrium cycling: for a maximum charge transferred of 50 mAh/g-NiHCF (mass NiHCF in single electrode), each electrode will change its equilibrium potential by approximately 0.2 V, and thus in the ideal case, without ion or electron transfer limitations, the maximum cell voltage is also 0.2 V. Instead, for the cycle with the highest charge, we observe the cell voltage approaching 1.0 to 1.5 V at the end of a half-cycle. Thus, transport limitations must have played a role in limiting the rate capability of the present CID experiments, as a result of sluggish pore-scale or solid-state transport processes. We reiterate that the charging behavior in 1 M $Na_2SO_4$ with higher current densities, as discussed in Fig. 2, did not show such strong rate limitations, but note that in such an electrolyte the Na-concentration is one-hundred times larger than in the CID experiments. This aspect of our electrode design requires further study and optimization.

### 3.4 Desalination Performance Metrics

We calculated the salt adsorption capacity (SAC), current efficiency, and salt-specific energy consumption for each of the cases reported in Fig. 3 at a current density of 2.8 A/m$^2$ and at 1.4 A/m$^2$. Figure 4 shows SAC and current efficiency as a function of electrode charge. SAC is expressed in mg NaCl adsorption per total mass of both electrodes (not per mass of active component, which in our case is 80 wt% of the total electrode mass). The highest value of SAC here, 34 mg/g, is about 2.5 times a reference value for CDI with carbon electrodes of ~12.5 mg/g [1,51]. Though SAC in CDI with carbons can be increased to higher values when using higher cell voltages, membranes, or chemically modified electrodes [52,53], on this metric CID using NiHCF IHCs competes quite well with carbon-based CDI technology. Note that in CID the value of SAC reported is reached in a half-cycle (thus in a full cycle, SAC is twice larger), whereas in a classical CDI cell a full cycle is necessary to accomplish the reported values of SAC. Fig. 4A also presents a limited number of data at a twice lower current density (1.4 A/m$^2$), related to one voltage-time trace that is shown in Fig. 5A. For both current densities, Fig 4B presents data for current efficiency $\lambda$, which is the ratio of average salt adsorption rate, $J_{salt,avg}$ (which is SAC multiplied by two times the electrode mass, $M_{el}$, divided by half-cycle time, HCT, and NaCl mass, $M_w$=58.44 g/mol), over current, $I$, in A, divided by $F$=96485 C/mol, and thus is given by

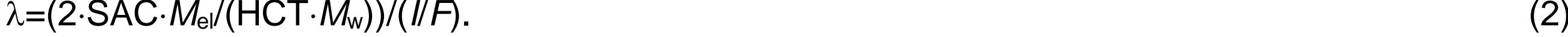

$$\lambda=(2\cdot SAC\cdot M_{el}/(HCT\cdot M_w))/(I/F). \quad (2)$$

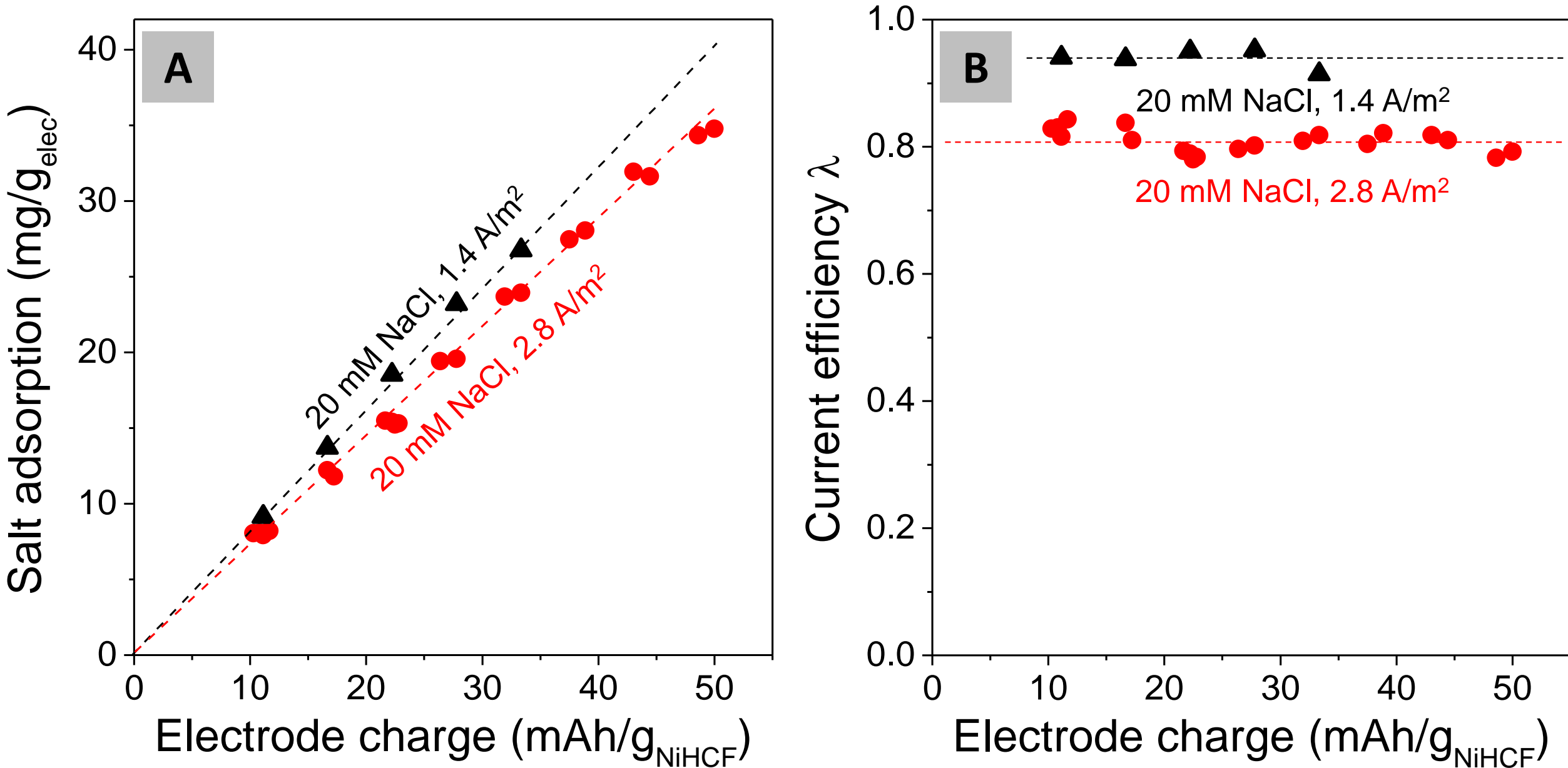

**Figure 4.** A) Salt adsorption capacity of NiHCF IHC electrodes, in mg NaCl per cycle per gram of both electrodes vs. electrode charge (per gram NiHCF in one electrode) for the experiments of Fig. 3 together with data at a 2x lower current density. B) Current efficiency, which is the ratio of salt adsorption rate over current density, see Eq. (2) ($\lambda$~0.8 for $I$=2.8 A/m$^2$, and $\lambda$~0.95 for $I$=1.4 A/m$^2$).

For current efficiency, $\lambda$, both data sets show a nearly perfect independence of $\lambda$ on electrode charge, with $\lambda$~0.8 at 2.8 A/m$^2$ and $\lambda$~0.95 at 1.4 A/m$^2$, see Fig. 4. Ideally, for perfect membranes (i.e., only counterion transport is permitted) and perfect cation intercalation, the current efficiency must be unity. Various factors may be responsible for the value of $\lambda$ being less than unity. One factor contributing to this effect, at least in part, is the loss of charge through undesired side reactions, such as $O_2$ evolution. Such a reaction is expected to occur to a greater degree when cycling with large overpotentials that are experienced at higher current densities. Furthermore, it is possible that NiHCF not only intercalates $Na^+$ ions, but also $Cl^-$ ions, though previous studies in K-based electrolytes concluded that cation intercalation in NiHCF predominates [54]. Based on a simple model we argue that salt back-diffusion through the AEM seems an unlikely explanation under the present experimental conditions (see SI). In conclusion, the influence of current density on current efficiency deserves further attention and the CID process and materials should be improved to increase $\lambda$ to a value much closer to unity, also at higher currents than presently employed.

Next, we analyze the energy invested in desalination. We reiterate that, for the present device using identical NiHCF electrodes, cell voltage (measured as the difference between the current collector potentials for anode and cathode as in Fig. 1, i.e., $V_{cell} = \phi_{s,a} - \phi_{s,c}$) takes equal and opposite values on average during charge and discharge, in contrast with previous reports of asymmetric electrochemical desalination devices [7,8,21] that show net positive cell voltage during both charge and discharge. For CID with symmetric electrodes energy recovery can affect energy consumption when ohmic losses are small (as predicted in Ref. [9]), but for the present experimental CID device the potential for energy recovery is negligible. In any case, in symmetric CID we can simply integrate cell voltage versus charge in one half-cycle to obtain energy consumption, i.e., cyclic integration (as in Refs. [8,21]) is unnecessary. Performing this analysis for the data at 17.6 mAh/g-NiHCF from Fig. 3B, which results in a SAC of around 12.5 mg/g (similar to many data for CDI with carbons), we obtain for 2.8 A/m$^2$ current density a value of ~60 kJ/mol salt removed and even as low as ~15 kJ/mol salt removed at a current of 1.4 A/m$^2$. These numbers are clearly below those of membrane CDI (using carbon electrodes with ion exchange membranes to increase current efficiency) excluding energy recovery and with brackish feedwater reported in Refs. [1,3]. This level of energy consumption also compares favorably with that of brackish water desalination using electrodialysis, where the minimum energy consumption for ~83% salt removal from 34 mM NaCl was reported as 47 kJ/mol-salt [55]. Experiments on the DB and a stationary CID cell in seawater (with 20 to 30 times higher ionic conductivity than the brackish solutions tested presently) show higher energy consumption than reported here when energy recovery is excluded (~ 25 kJ/mol-salt [8,21]), while these devices show lower energy consumption including energy recovery (8.5 kJ/mol-salt [8] and 6.4 kJ/mol-salt [21], respectively). That said, the present energy consumption values are of an illustrative nature and proper comparison can only be made for the same salt removal rate (in mol/m$^2$-s) and raw salt removal (in mM), where the present CID device shows lower values of both. There is also substantial opportunity for improving the cycling rate with better designed electrodes, as NiHCF electrodes have been cycled with 80% theoretical capacity with discharge times as fast as 10 min [28]. These preliminary results reveal the potential for reduced energy consumption using intercalation-based electrodes for electrochemical desalination. We note that for the 0.5 mM concentration change produced in the present CID experiments the theoretical minimum energy consumption is 0.13 kJ/mol-NaCl. Comparison with the present energy consumption levels suggests that dissipative energy losses are currently significant. We note, though, that similar NiHCF materials in aqueous batteries have produced discharge times of approximately 1 min with over 60% of theoretical capacity [28].

Thus, future work will address reducing dissipative energy losses. We can conclude from these findings and from Fig. 4 that there is potential for enhancement in salt adsorption capacity and a reduction in energy use of electrochemical water desalination when employing intercalation electrodes.

### 3.5 Desalination Dynamics with KCl/NaCl Mixtures

Finally, we present data for experiments with mixtures of NaCl and KCl, as summarized in Fig. 5 and Table S1. We note that the thermodynamics of adsorption of $Na^+$ and $K^+$ in NiHCF-electrodes (containing binder, carbon and IHC) from aqueous NaCl/KCl mixtures at pH 2 has also been explored recently by galvanostatic cycling and cyclic voltammetry [10]. Equilibrium adsorption analysis (based on both ideal solid-solution models [10] and, here, based on an interacting solid-solution model, as described in SI), indicates a substantial preference for $K^+$ adsorption over $Na^+$ that produces a greater than one-hundred fold selectivitiy ratio. In a CID device cycled at finite rate, non-equilibrium conditions are likely to prevail, and selectivity may be governed by factors other than thermodynamics. Our results for charging and discharging of a CID cell with NiHCF electrodes show preferential adsorption of $K^+$ over $Na^+$, as evidenced by the measured salt effluent concentration profile, obtained by an on-line measurement (see Methods), but selectivity is substantially less than expected based on thermodynamics. For the experiments with a 1:1 $K^+$:$Na^+$ concentration ratio in the inflow solution, we observe more $K^+$-adsorption than adsorption of $Na^+$, while for the experiment with the $K^+$:$Na^+$ inflow ratio three times reduced, the adsorption of $Na^+$ and $K^+$ are about the same. Thus, in both cases the separation factor, defined as [34,36]

$$\alpha_{K:Na}=(\Delta[K^+]_{ads}/\Delta[Na^+]_{ads})/([K^+]_{sol}/[Na^+]_{sol}) \qquad (3)$$

is above $\alpha_{K:Na}=3$. The definition of $\alpha$ relates the (ratio of) the concentration *change* of $K^+$ and $Na^+$ in the IHC (linearly dependent on adsorbed amount, expressed in mmol/g, see Table S1) to the ratio of $K^+$:$Na^+$ concentrations in solution (which we base on the feed concentrations). This value of the separation factor, $\alpha_{K:Na}$, is very comparable in magnitude to values reported in ref. [34]. That our values are about a factor 2 lower than in ref. [34] must be due to the fact that in our experiment, thermodynamic equilibrium was not reached. Interestingly these values for $\alpha$ are clearly less than the value of $\alpha$>100 predicted theoretically based on the equilibrium titration curve of $K_2SO_4$ and $Na_2SO_4$ (see Fig. S5 in SI and ref. [10]). In Fig. 5 and Table S1 we compare with one data point for $K^+$:$Na^+$ mixtures tested with activated carbon electrodes in CDI by Dykstra *et al.* [43], which showed no preferential adsorption for $K^+$ over $Na^+$, in line with the adsorption mechanism based on capacitive double layer formation. Though the CID concept was previously predicted for use in desalinating NaCl [9], based on the present data we can conclude that CID with NiHCF intercalation host compounds as active phase in mixed porous electrodes provides much potential for the selective removal of metal ions from water.

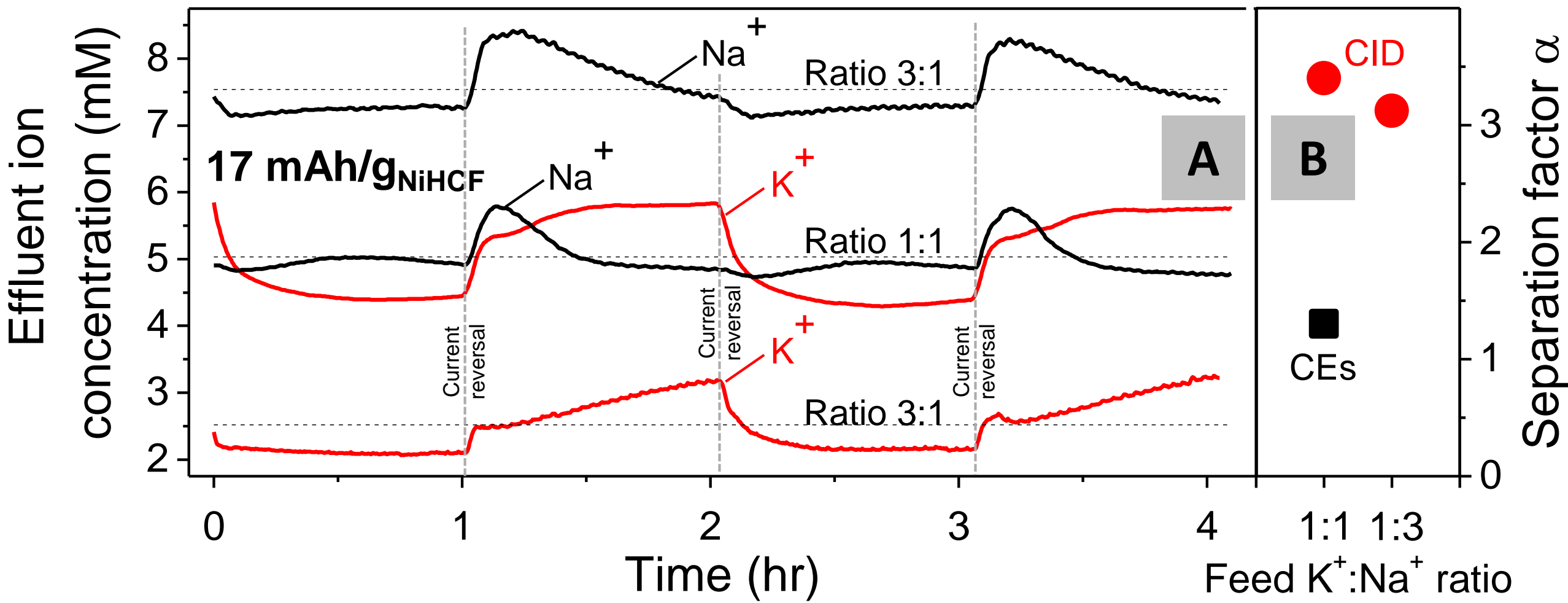

**Figure 5.** Desalination from $K^+$:$Na^+$ mixtures at two feed mixing ratios (current *I*=2.8 $Am^2$). Left: traces of effluent ion concentration in two consecutive charging/discharge cycles; right: data for the resulting separation factor $\alpha_{K:Na}$, compared with one data point for CDI using carbon electrodes (CEs).

**Conclusions**

We have shown experimentally that cation intercalation desalination (CID) cells using nickel hexacyanoferrate intercalation electrodes, separated by an anion exchange membrane, can be used to desalinate NaCl aqueous solutions. In so doing, we find that salt adsorption capacity is increased at various levels of current and cycle duration. These results constitute the first experimental demonstration of CID, a concept that was previously theorized [9,19]. As a result, this study using nickel hexacyanoferrate electrodes provides a benchmark for the use of other intercalation host compounds and cells designs in CID. We also show that, when nickel hexacyanoferrate is employed in CID, $K^+$ ions can be removed selectively from mixtures with $Na^+$ ions, and further investigation is needed to understand the role of transport and thermodynamic effects in such cells when water sources containing mixtures of cations are desalinated. Presently, we consider desalination from brackish waters, but applications with other water sources (e.g., sea and waste water) are also of interest. Furthermore, the degree and the rate of salt removal tested here are sufficient to demonstrate the CID concept, but further study is needed to evaluate the impact of these effects on energy consumption.

## Acknowledgments

KCS and AS acknowledge support from the Department of Mechanical Science and Engineering and the College of Engineering at the University of Illinois at Urbana-Champaign. Part of this work was performed in the cooperation framework of Wetsus, European Centre of Excellence for Sustainable Water Technology (www.wetsus.eu). Wetsus is co-funded by the Dutch Ministry of Economic Affairs and Ministry of Infrastructure and Environment, the Province of Fryslân, and the Northern Netherlands Provinces.

# Supplementary Information

S. Porada, A. Shrivastava, P. Bukowska, P.M. Biesheuvel, and K.C. Smith

## 1 Additional NiHCF characterization

Scanning electron microscopy (SEM) was used to obtain images of the nanoparticle aggregates prior to ball milling and the porous electrodes containing the PBA compound (Fig. S1). Elemental analysis was conducted with NiHCF decomposed into elemental species in acid using a PerkinElmer 2400 Series II CHN/O Elemental Analyzer and a PerkinElmer 2000DV ICP-OES instrument. X-ray diffraction was conducted using a Bruker D8 Venture instrument with Cu-K$\alpha$ radiation ($\lambda$ = 1.54 Å, 0.02° angular resolution, and 3° < $2\theta$ < 100°).

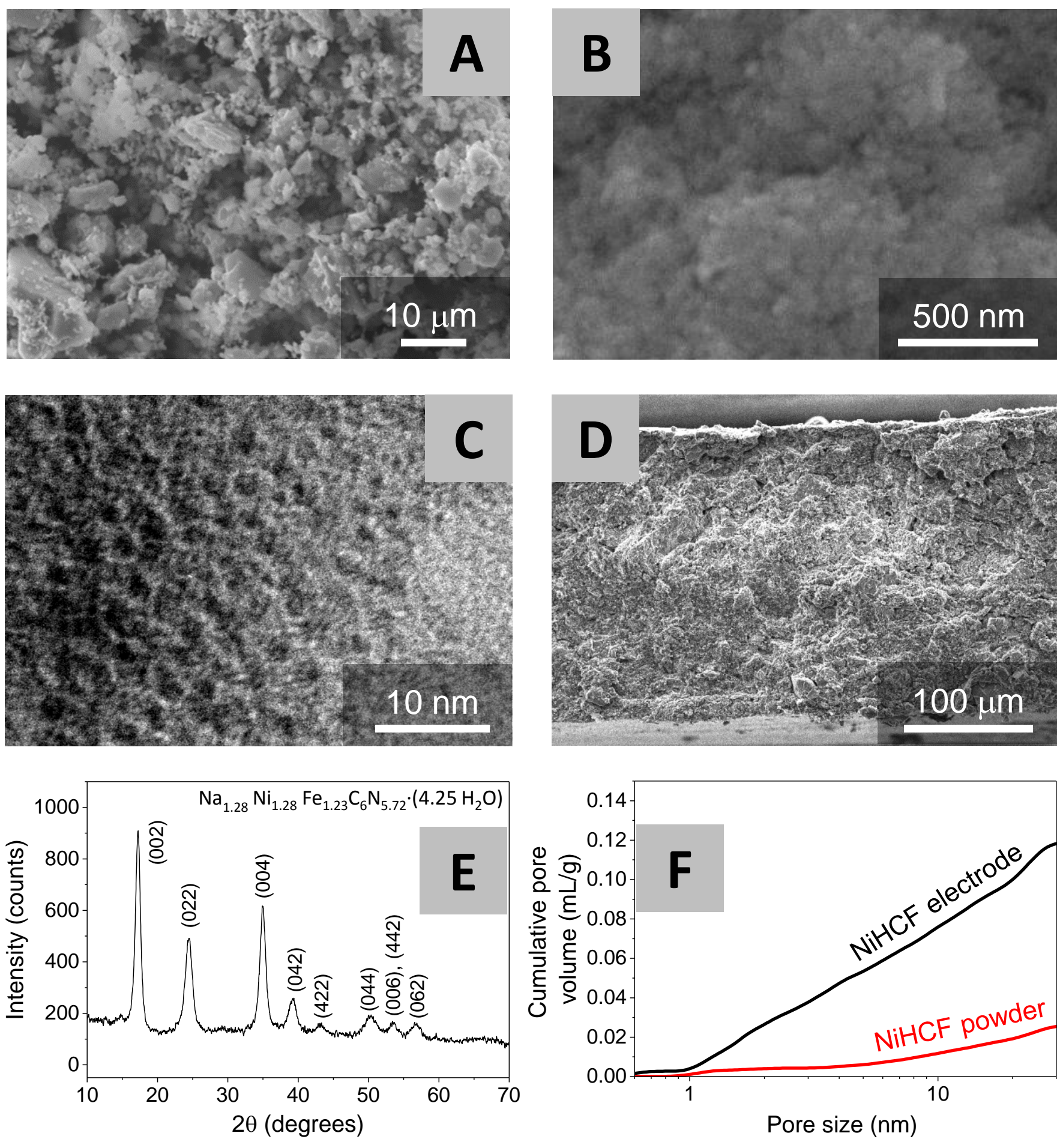

Figure S1: (A, B) Scanning electron and (C) transmission electron images of the NiHCF powder. (D) Cross-section scanning electron image of the final NiHCF-electrode, and (E) X-ray diffraction diagram for the as-synthesized NiHCF powder. (F) Cumulative pore size distribution of the NiHCF powder and of the final NiHCF-electrode.

## 2 Differential capacitance of NiHCF electrodes

Based on the voltage-charge of a single NiHCF electrode, as presented in Fig. 2B in the main text, we can construct a graph of differential capacitance, $C_d$, which is the local gradient of electrode charge vs electrode voltage. Like the data for charge in Fig. 2B, these data for $C_d$ are presented per gram of NiHCF in a single electrode. In Fig. S2A, we present these data versus electrode charge, where charge is shifted such that charge=0 corresponds to the optimum in the curve (vertical symmetry line). The theory-line is the same Frumkin-equation as used in Fig. 2B in the main text. In Fig. S2, we compare with typical values for the differential capacitance of activated carbon (AC) electrodes, around 80 F/g (per gram of activated carbon), which are quite independent of charge, in line with data e.g. reported in ref [1]. For AC, a maximum electrode charge of ~ 14 mAh/g corresponds to ~ 50 C/g (pure AC) which is reached at a cell voltage of ~ 1.2 V, a natural maximum for CDI with ACs to avoid water splitting. The charge-range in Fig. S2 for NiHCF seems to be less than a factor two higher than for AC, but note that for NiHCF as used in CID, the entire range is available in a charge-discharge process (from almost full deintercalation on one extreme to full intercalation on the other extreme), while for most CDI processes only half of the range (green horizontal bar) is used, because discharge is at zero charge. Thus, in this comparison, the total charge that can be transferred per gram of active component is about a factor of three higher for NiHCF than for AC. Fig. S2 shows that in a large range of charge, the differential capacitance of NiHCF is fairly constant, only to drop off significantly when we reach full (de)intercalation. Interestingly, if we plot the differential capacitance versus electrode voltage, see Fig. S2B,we observe a very strong dependence on electrode potential, with the entire data set confined to a narrow range of 0.2 V of electrode potential, and $C_d$ constant in a voltage window of only 0.1 V. This narrow range of potential does not imply that the differential capacitance is not constant over a large range of process conditions (charging degree of the NiHCF), for which inspection of Fig. S2A instead shows that over a range of ~ 40 mAh/g, the differential capacitance $C_d$ of NiHCF electrodes is fairly constant.

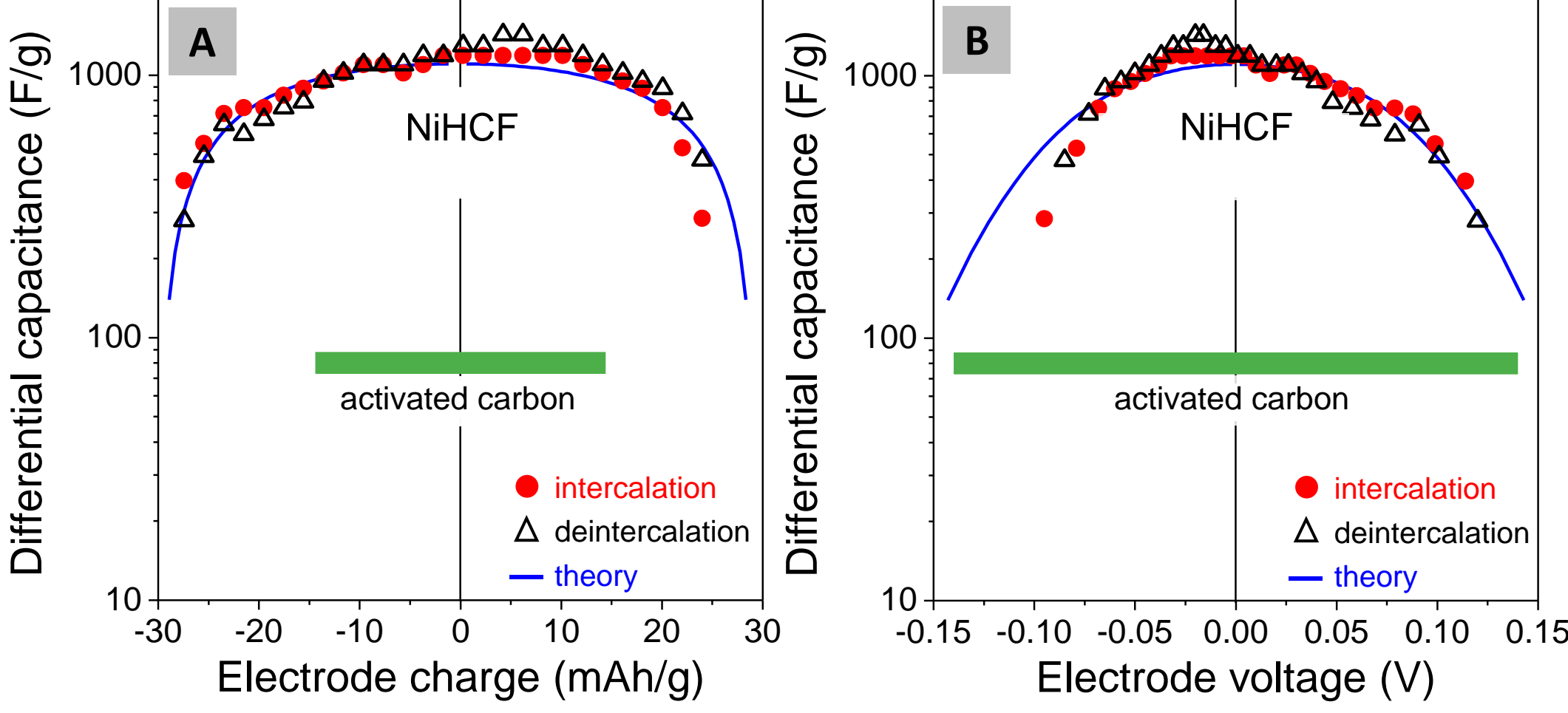

Figure S2: Differential capacitance of NiHCF-electrodes based on data in Fig. 2B, in comparison with activated carbon electrodes, as function of A) electrode charge, and B) electrode potential. x-Axis values for charge and potential are shifted such that a value of zero corresponds to a maximum capacitance (symmetry point).

# 3 Additional Desalination Experiments in NaCl

Results of two sets of experiments complementary to those presented in Fig. 3 for pure NaCl-solutions were performed and will be discussed next. First, at the same current density as Fig. 3 (2.8 A/m$^2$), the two electrodes in the CID cell were initially intercalated to approx. 40-60% of their full capacity. In these experiments half-cycle time was increased after each 2-3 cycles, rather than decreased. The resulting time evolution of effluent salinity and cell voltage is shown in Fig. S2. Results are quite similar to those in Fig. 3 of the main text.

In addition, experiments were done at a two-times reduced current density (~ 1.4 A/m$^2$), see Fig. S4. In this experiment the change of salt concentration is reduced by a factor two compared to the experiment at the higher current, while the cell voltage is reduced by a larger factor, as discussed in the main text. In this experiment, we go from short cycles to long cycles, like in Fig. S3.

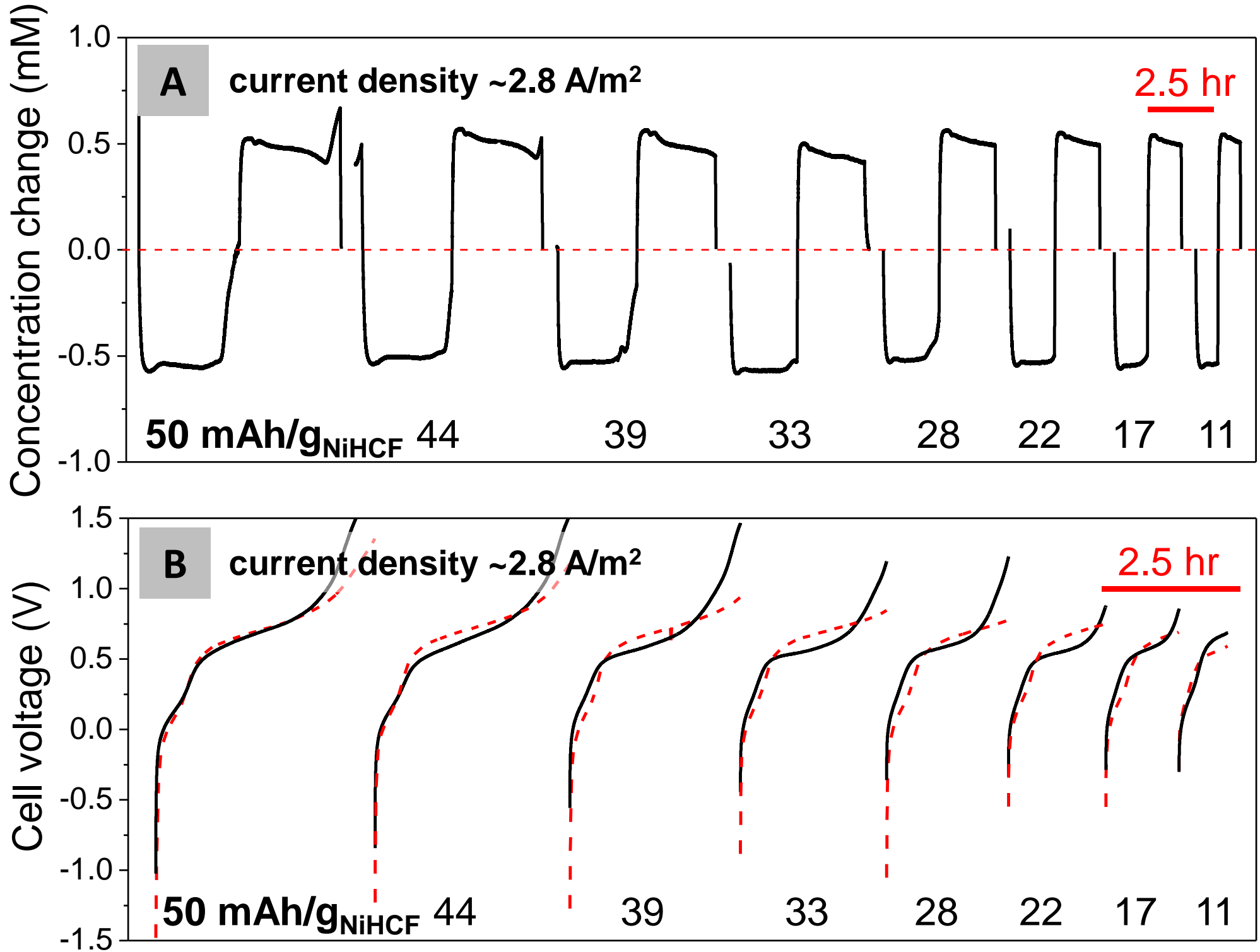

Figure S3: Desalination cycles using NiHCF IHC electrodes at a constant current of 2.8 A/m$^2$ in 20 mM NaCl, for sets of cycles where the charging time is increased after each 2 or 3 cycles, starting at a situation where both electrodes are initially intercalated to a similar degree of $\theta$ between 0.4 to 0.6. In other words elapsed time proceeds from right to left in the figures. Panel A presents one cycle from each set of 2 or 3 similar cycles, showing both desalination (concentration change < 0) and salt release (concentration change > 0). Note that here only results for the effluent salinity from one of the two channels is shown. In panel B profiles of cell voltage versus time are shown during the same desalination cycle. Here, to show the symmetry in the process, the voltage trace during the second half of the cycle is vertically switched and moved left, to overlap with the voltage trace in the earlier half-cycle.

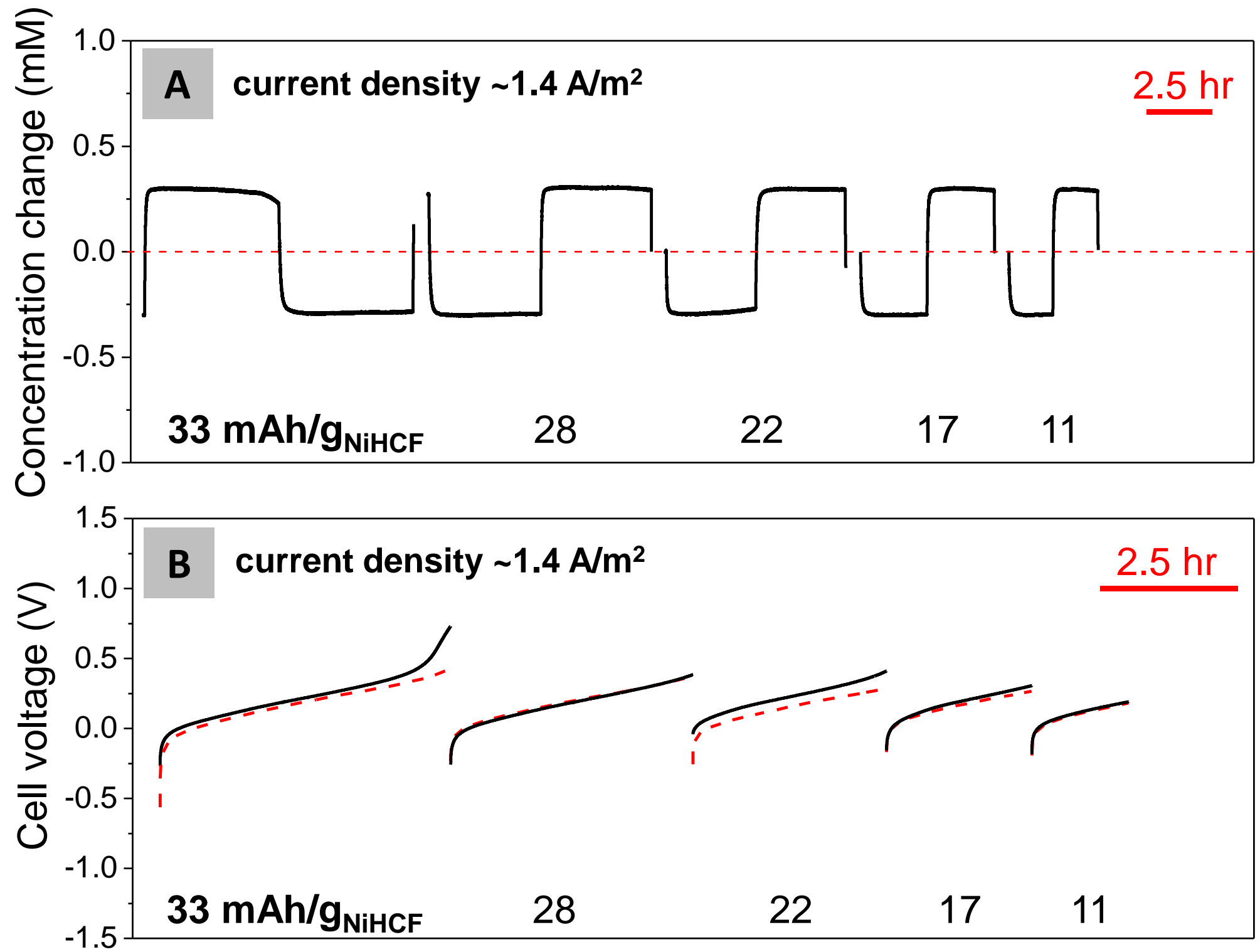

Figure S4: Desalination cycles using NiHCF IHCs at a constant current of 1.4 A/m$^2$ in 20 mM NaCl, for sets of cycles where the charging time is increased from cycle to cycle, like Fig. S3. See Fig S3 for explanation of each panel.

# 4 Additional Half-Cell Characterization Experiments

To support the experiments with NaCl/KCl mixtures, we also performed GITT titrations (like reported in Fig. 2) both in $K_2SO_4$ and $Na_2SO_4$. Because $K_2SO_4$ has a solubility below 1 M, we decided for salt concentrations of 0.1 M (Na-ion and K-ion concentrations are twice higher than that). At this lower salinity, in this experiment, we are not able to achieve the maximum charge of 59 mAh/g (per gram NiHCF) as in Fig. 2, but about 10% less. For $Na_2SO_4$, Eq. (1) from the main text describes the data, with the parameter settings for $E_{\mathrm{Na,ref}}$ and $g_{\mathrm{Na}}$ as derived for testing in 1 M $Na_2SO_4$, i.e., the theoretical curve in Fig. S3 is shifted downward by $V_T \ln(10) \sim 59$ mV relative to that in Fig. 2. For $K_2SO_4$, as expected based on literature, the data are at a higher voltage, and in addition, we find a less horizontal plateau, i.e., the capacitance (derivative of charge with voltage) is lower, namely approx. 800 F/g between an electrode charge of 15 and 45 mAh/g. We fit the data for $K_2SO_4$ with $E_{\mathrm{K,ref}} = 555$ mV and $g_{\mathrm{K}} = 150$ mV. The experimental data for titration of NiHCF in 0.5 M $K_2SO_4$ reported by Erinmwingbovo *et al.* [2] give a much higher capacitance, namely for intercalation degrees in between the 0.4 to 0.7 we calculate a value of more than 1400 F/g (based on a maximum charge of 59 mAh/g), which is also larger than what we obtain for 1 M $Na_2SO_4$ (Fig. 2).

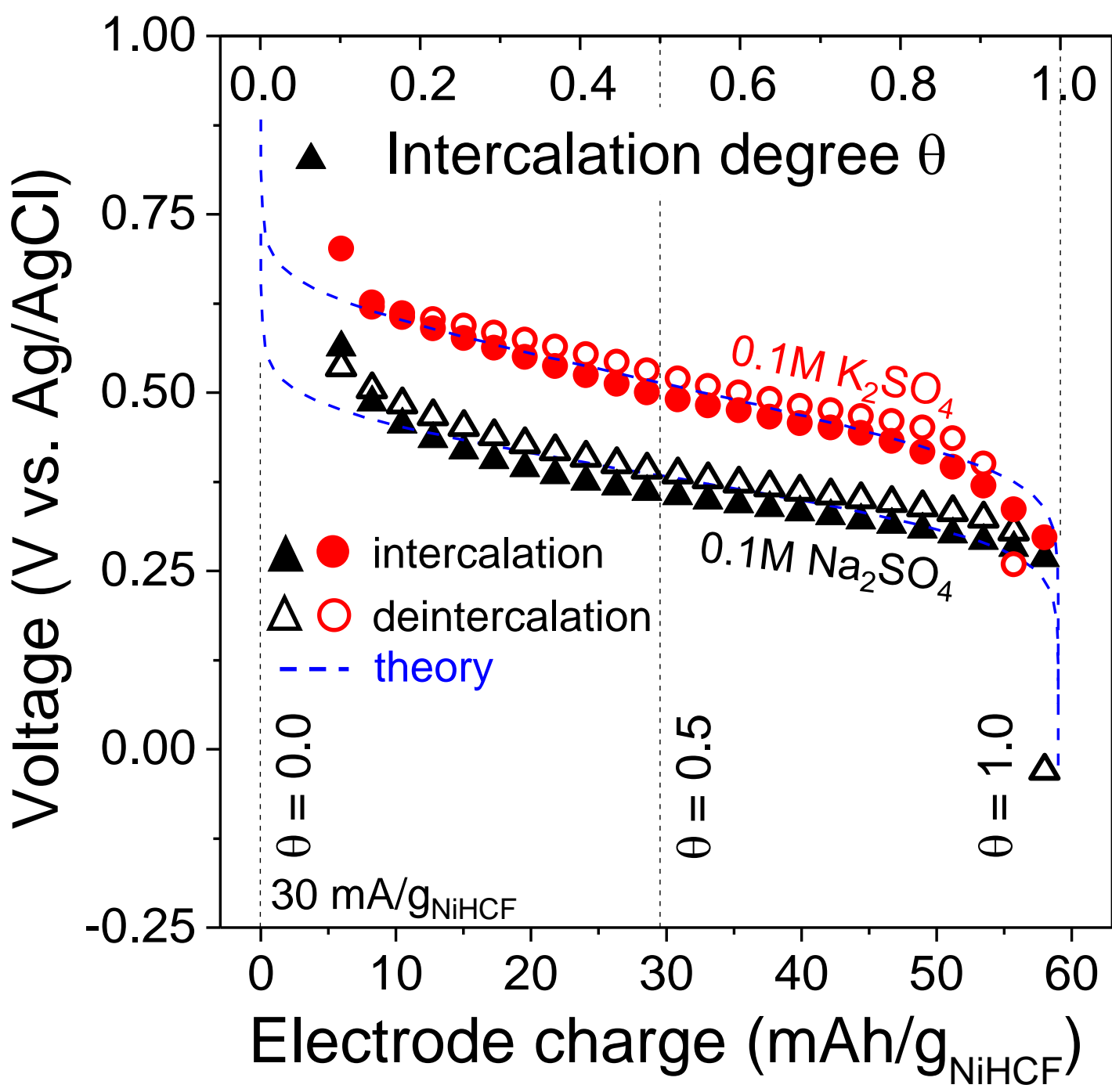

Figure S5: Electrochemical titration by GITT of an NiHCF electrode in 0.1 M $Na_2SO_4$ (triangles) and 0.1 M $K_2SO_4$ (circles). In this experiment, a constant current of 30 mA/g of NiHCF is applied for 270 s, after which current is set to zero for 60 s. At the end of this period, we log a charge-voltage data point. The theory lines were obtained based on the Frumkin-equation, see Eq. (1) in the main text.

# 5 Desalination experiments in NaCl/KCl-mixtures

We desalinated water containing a mixture of NaCl and KCl in our CID-cell with NiHCF, as reported in Fig. 6 in the main text (current $I = 2.77$ A/m$^2$; water flow rate per channel 4.7 mL/min). We summarize various essential data of the two experiments (at two different NaCl/KCl mixing ratios) in Table S1, and compare with one CID-experiment at 20 mM NaCl (as reported in Figs. 3, 4 and 5 of the main text) and with one CDI-experiment with carbon electrodes, from ref. [3].

The CID experiments are done for a total charge transfer of 17 mAh/g in one (dis-)charging step (half-cycle), after which current is reversed and the same charge is again transferred, but now in the other direction. The cation adsorption, $\Gamma_{i,\mathrm{ads}}$ ($5^{\mathrm{th}}$ and $6^{\mathrm{th}}$ column in Table S1), is the cation removal from one channel during a half-cycle. These removal amounts are defined per gram of NiHCF in a single electrode. We can multiply by the mass density of the NiHCF of $\rho = 2.0$ g/mL, to obtain the concentration *change* for the cation in the active material between the start and end of each half-cycle, $\Delta[\mathrm{C}^+]_{\mathrm{ads}}$. Applying Eq. (3) from the main text, we can calculate the separation factor $\alpha_{\mathrm{K:Na}}$. Note that in this experiment, equilibrium was not reached because of constant-current operation. To reach equilibrium, the cell voltage should be kept at a certain value for a prolonged time, before the current is reversed.

| | Feed $[\mathrm{K}^+]$:$[\mathrm{Na}^+]$ ratio | $[\mathrm{K}^+]_{\mathrm{sol}}$ (mM) | $[\mathrm{Na}^+]_{\mathrm{sol}}$ (mM) | $\Gamma_{\mathrm{K+,ads}}$ (mmol/g) | $\Gamma_{\mathrm{Na+,ads}}$ (mmol/g) | $\Delta[\mathrm{K}^+]_{\mathrm{ads}}$ (mM) | $\Delta[\mathrm{Na}^+]_{\mathrm{ads}}$ (mM) | $\alpha_{\mathrm{K:Na}}$ | |
|---|---|---|---|---|---|---|---|---|---|
| | 0:1 | 0 | 20 | | 0.520 | | | | |
| CID | 1:1 | 5.0 | 5.0 | 0.268 | 0.079 | 540 | 160 | 3.4 | @ 17 mAh/g |
| | 1:3 | 2.5 | 7.5 | 0.150 | 0.155 | 300 | 310 | 2.9 | |
| CE | 1:1 | 5.0 | 5.0 | 0.120 | 0.100 | 200 | 165 | 1.2 | charge/discharge @1.0/0 V |

Table S1: Experimental results for CID with $\mathrm{K}^+$:$\mathrm{Na}^+$ mixtures and the calculation of the separation factor $\alpha_{\mathrm{K:Na}}$. Data are compared with one experiment for CDI with carbon electrodes (CE). For CID, NiHCF density $\rho = 2.0$ g/mL; for carbon electrodes, micropore volume $v_{\mathrm{mi}} = 0.6$ mL/g.

The CDI experiment with carbon electrodes from ref. [3] is based on a mixture of 5 mM NaCl and 5 mM KCl. In the experiment salt adsorption by the pair of electrodes is measured in mol/g. Assuming only cation adsorption in the cathode (and vice-versa, anion adsorption in the anode), we come to the numbers for $\Gamma_{i,\mathrm{ads}}$ in Table S1. Multiplying by micropore volume $v_{\mathrm{mi}}$ results in predictions for the concentration change for $K^+$ and $Na^+$ in the cathode micropores.

# 6 Desalination experiments with both effluent salinities measured

In separate CID-experiments with 20 mM NaCl solution, we measured the salt effluent concentration of the solution coming from both channels. As expected, the two patterns are very similar, with concentrations going up and down in an anti-parallel manner. All experiments are performed at 10 mA current (2.77 A/m$^2$), for the first four cycles for a duration per half-cycle of 1000 s (17 min). For the second set of two cycles, the half-cycle time is 2000 s (34 min). The water flow rate through each channel is 7.5 mL/min. The electrode is manufactured from a recipe containing 20 wt% carbon black (not 10% as for the other results), 10 wt% binder and 70 wt% NiHCF. The cell is constructed with the flow channel not behind the electrode as in the main text, but located in front, i.e., the flow channel is in between membrane and electrode.

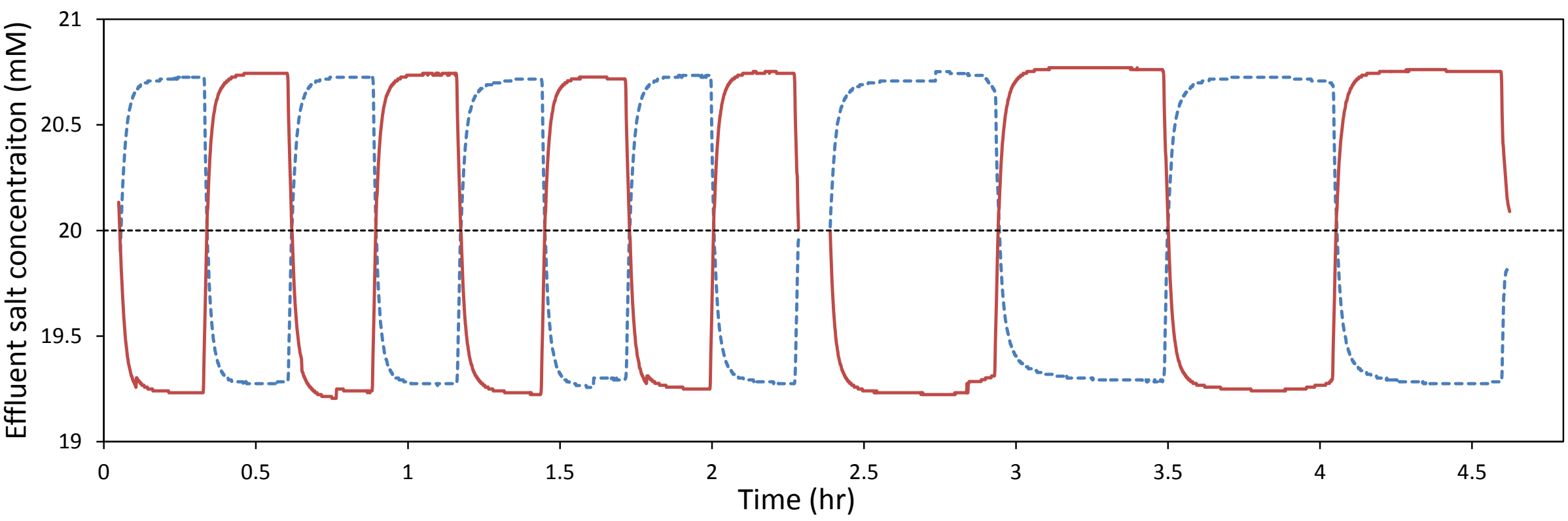

Figure S6: CID-experiment with NaCl solution where the salinity of the water coming from both channels is simultaneously measured, based on measuring conductivity. The cell is constructed with a flow channel in between membrane and electrode. Other parameter settings in main text.

# 7 Thermodynamics of Intercalation host compounds in salt mixtures

The Frumkin equation for a salt solution with only one cation (say, $Na^+$) is given by

$$E = E_{i,\mathrm{ref}} - V_{\mathrm{T}} \left( \ln \frac{\theta_i}{1-\theta_i} - \ln \frac{c_{i,\infty}}{c_0} \right) - g_i \left( \theta_i - 1/2 \right) \tag{1}$$

where $E$ is the electrode potential, $E_{i,\mathrm{ref}}$ is an ion-dependent potential which relates to the affinity of the ion with the IHC matrix, $V_{\mathrm{T}}$ is the thermal voltage (~ 25.6 mV at room temperature), subscript $\infty$ refers to ion concentrations in solution outside the IHC, and $c_0$ is a reference concentration, $c_0$ = 1 M. Intercalation degree, $\theta_i$, is a parameter which runs between 0 and 1 when the material goes from the maximum (practically achievable) de-intercalated state, to the state of full intercalation, see Fig. 2B of main text.

Eq. (1) can be derived from a balance of chemical equilibrium of an ion in equilibrium between solution and the IHC, where the various terms can be derived from analysis of (separate contributions to) the free energy density, $f$. For the last term of Eq. (1) which relates to ion-ion interaction in the material we can make the following derivation based on $f$ for mixtures.

For a salt mixture containing components “i” and “j”, the free energy density for the mixing enthalpy is given by

$$f \propto \frac{1}{2} g_i \theta_i^2 + \frac{1}{2} g_j \theta_j^2 + g_{\mathrm{avg}} \theta_i \theta_j . \tag{2}$$

Via the chemical potential of species i, $\mu_i \propto \partial f/\partial\theta_i$, we can derive the modified Frumkin-equation for binary mixtures,

$$E = E_{i,\mathrm{ref}} - V_\mathrm{T}\left(\ln\frac{\theta_i}{1-\theta_i-\theta_j} - \ln\frac{c_{i,\infty}}{c_0}\right) - g_i\left(\theta_i - 1/2\right) - g_\mathrm{avg}\theta_j \tag{3}$$

where for $g_\mathrm{avg}$ we take the average of $g_i$ and $g_j$. For a given electrode potential $E$, Eq. (3) must be set up twice, for ion i and j, and solved jointly, to obtain the individual intercalation degrees $\theta_i$ and $\theta_j$.

In Fig. S3 we present equilibrium titration data by GITT (like Fig. 2B) at 0.1 M $Na_2SO_4$ and 0.1 M $K_2SO_4$. The theory lines are based on Eq. (1) from the main text, with for $Na_2SO_4$ the values for $E_\mathrm{Na,ref}$ and $g_\mathrm{Na}$ given in the main text ($E_\mathrm{Na,ref}$ = 425 mV, $g_\mathrm{Na}$ = 90 mV), while for $K_2SO_4$ we find a good fit using $E_\mathrm{K,ref}$ = 555 mV and $g_\mathrm{K}$ = 150 mV. Using these values for $E_{i,\mathrm{ref}}$ and $g_i$ (with $g_\mathrm{avg}$ = 120 mV), we find that in the typical potential range, according to thermodynamic equilibrium the IHC is predominantly filled with $K^+$, with (for a solution of 5 mM KCl and 5 mM NaCl) separation factors $\alpha_\mathrm{K:Na} > 160$ for $E > 200$ mV.

# 8 Salt Backdiffusion in Ion-exchange Membranes

The ion flux across a membrane is given by the Nernst-Planck equation [4],

$$J_i = -D_i\left(\frac{\partial c_i}{\partial x} + z_i c_i \frac{\partial\phi}{\partial x}\right). \tag{4}$$

For a 1:1 salt, local membrane electroneutrality is

$$c_+ - c_- + \omega X = 0 \tag{5}$$

where $\omega$ is the sign of membrane immobile charge, $\omega = +1$ for an anion-exchange membrane, and where $X$ is the value of the fixed membrane charge (per unit open volume), typically around 5 M for a commercial membrane. Combining Eq. (4) with Eq. (5), assuming equal ion diffusioen coefficients, leads to

$$J_\mathrm{ions} = J_+ + J_- = -D\left(\frac{\partial c_\mathrm{T}}{\partial x} - \omega X\frac{\partial\phi}{\partial x}\right) \qquad \text{and} \qquad J_\mathrm{ch} = J_+ - J_- = -D c_\mathrm{T}\frac{\partial\phi}{\partial x} \tag{6}$$

where $c_\mathrm{T}$ is the total ions concentration in the membrane. In steady-state, fluxes are invariant with $x$, and we can integrate both equations across the membrane, assuming a linear profile for the inverse of $c_\mathrm{T}$, to arrive at

$$J_\mathrm{ions} = -\frac{D}{\ell}\left(c_\mathrm{T,R} - c_\mathrm{T,L} - \omega X\Delta\phi\right) \tag{7}$$

and

$$J_\mathrm{ch} = -\frac{D}{\ell}\langle c_\mathrm{T}\rangle\Delta\phi \tag{8}$$

where

$$\frac{2}{\langle c_\mathrm{T}\rangle} = \frac{1}{c_\mathrm{T,L}} + \frac{1}{c_\mathrm{T,R}} \tag{9}$$

where "L" and "R" refer to the left and right side of the membrane. The current efficiency, $\lambda = \frac{J_\mathrm{ions}}{J_\mathrm{ch}}$, which in a steady state process has the same value as $\Lambda$, the charge efficiency, thus $\lambda \sim \Lambda$, is given by combination of Eqs. (7) and (8), resulting in

$$\lambda = -\frac{D}{J_\mathrm{ch}\ell}\left(c_\mathrm{T,R} - c_\mathrm{T,L}\right) - \frac{\omega X}{\langle c_\mathrm{T}\rangle} \tag{10}$$

where the ions concentration at the "L" and "R"-edges is related to the salt concentration just outside the membrane, $c_\infty$, by

$$c_\mathrm{T}^2 = X^2 + \left(2c_\infty\right)^2. \tag{11}$$

A current density of 2.8 A/m$^2$ results in $J_\mathrm{ch}$ = 28 $\mu$mol/m$^2$/s. For the Neosepta anion-exchange membrane we can use $X$ = 5 M, and $\ell \sim 130$ $\mu$m, while $D$ can be of the order of 10 % of the value in free solution, thus approx. $D \sim 10^{-10}$ m$^2$/s. The two concentrations on each side of the membrane are about ~ 0.5 mM higher and lower than the inlet value of 20 mM. With these parameter settings, $\lambda = 0.9995$, and thus the membrane is expected to be perfectly cation-blocking. Variation of these parameters, such as reducing the diffusion coefficient, leads to only marginal variations in $\lambda$.

In conclusion, the observed low value of $\Lambda$ is not due to membrane salt back-diffusion.